\documentclass[runningheads]{llncs}

\usepackage{amsmath}
\usepackage{amsfonts}
\usepackage{amssymb}
\usepackage{booktabs}
\usepackage{graphicx}
\usepackage[colorlinks]{hyperref}
\usepackage[switch]{lineno}
\usepackage[normalem]{ulem} 
\usepackage[all]{xy}
\usepackage{cancel}
\usepackage{multirow}

\graphicspath{{figures/}}

\newcommand{\appendixref}[1]{\hyperref[#1]{Appendix~\ref*{#1}}}


\DeclareMathOperator{\suchthat}{:}

\newcommand{\DND}{\textbf{DND}}
\newcommand{\DB}{\textbf{DB}}

\newcommand{\cand}{\mathcal{C}}

\newcommand{\stand}{\mathcal{S}}

\newcommand{\numcands}{k}

\newcommand{\altorder}{alt-order}
\newcommand{\Altorder}{Alt-order}

\newcommand{\AWAIREold}{{AWAIRE~v1}}
\newcommand{\AWAIREnew}{{AWAIRE~v2}}

\newcommand{\cw}{\texttt{W}}
\newcommand{\cx}{\texttt{X}}
\newcommand{\cy}{\texttt{Y}}
\newcommand{\cz}{\texttt{Z}}
\newcommand{\X}{\texttt{X}}
\newcommand{\Y}{\texttt{Y}}
\newcommand{\Z}{\texttt{Z}}
\newcommand{\W}{\texttt{W}}
\newcommand{\pp}{\oplus}

\title{Improving the Computational Efficiency of Adaptive Audits of IRV
Elections\thanks{We thank Vanessa Teague for helpful discussions.
This work was supported by the Australian Research Council
(Discovery Project DP220101012, OPTIMA ITTC IC200100009) and the US\ National
Science Foundation (SaTC~2228884).}}

\author{
Alexander Ek       \inst{1}  \orcidID{0000-0002-8744-4805}  \and
Michelle Blom      \inst{2}  \orcidID{0000-0002-0459-9917}  \and
Philip B. Stark    \inst{3}  \orcidID{0000-0002-3771-9604}  \and
Peter J. Stuckey   \inst{4}  \orcidID{0000-0003-2186-0459}  \and
Damjan Vukcevic    \inst{1}  \orcidID{0000-0001-7780-9586}
}

\authorrunning{Ek A, Blom M, Stark PB, Stuckey PJ, Vukcevic D}

\institute{
Department of Econometrics and Business Statistics, Monash University, Clayton,
Australia
\and
Department of Computing and Information Systems, University of Melboure,
Parkville, Australia
\and
Department of Statistics, University of California, Berkeley, CA, USA
\and
Department of Data Science and AI, Monash University, Clayton, Australia
\\
\email{damjan.vukcevic@monash.edu}
}

\begin{document}

\maketitle

\begin{abstract}
AWAIRE is one of two extant methods for conducting risk-limiting audits of
instant-runoff voting (IRV) elections.  In principle AWAIRE can audit IRV
contests with any number of candidates, but the original implementation
incurred memory and computation costs that grew superexponentially with the
number of candidates.  This paper improves the algorithmic implementation
of AWAIRE in three ways that make it practical to audit IRV contests with
55~candidates, compared to the previous 6~candidates.  First, rather than
trying from the start to rule out all candidate elimination orders that
produce a different winner, the algorithm starts by considering only the
final round, testing statistically whether each candidate could have won
that round.  For those candidates who cannot be ruled out at that stage, it
expands to consider earlier and earlier rounds until either it provides
strong evidence that the reported winner really won or a full hand count is
conducted, revealing who really won.  Second, it tests a richer collection
of conditions, some of which can rule out many elimination orders at once.
Third, it exploits relationships among those conditions, allowing it to
abandon testing those that are unlikely to help.  We provide real-world
examples with up to 36~candidates and synthetic examples with up to
55~candidates, showing how audit sample size depends on the margins and on
the tuning parameters.  An open-source Python implementation is publicly
available.
\end{abstract}


\section{Introduction}

\emph{Risk-limiting audits} (RLAs) are gaining attention in the world of
election security and assurance.\footnote{%
\url{https://www.ifes.org/publications/risk-limiting-audits-guide-global-use},
and \url{https://www.ncsl.org/elections-and-campaigns/risk-limiting-audits}
(visited 15~May 2024)}
An RLA is any procedure with a guaranteed minimum probability of correcting the
reported outcome of an election if the reported outcome is wrong, and that
never alters a correct outcome.\footnote{%
    The collection of ballot cards from which the audit sample is drawn must be
    a demonstrably complete and trustworthy record of the validly cast votes;
    otherwise, no audit procedure can guarantee a nonzero chance of catching
    and correcting wrong outcomes.  See, e.g., Appel \& Stark
    \cite{appelStark20}.}
\emph{Outcome} means who or what won, not the vote tallies.
The \emph{risk limit} $\alpha$ is the maximum chance that a wrong outcome will
not be corrected.  RLAs generally involve sampling cast ballot cards at random
and manually reading the votes on those cards.  RLAs can use a broad variety of
sampling designs and can use a variety of information from the voting system to
improve efficiency \cite{shangrla,stark2023alpha,stark23ONEAudit}.

Improving the efficiency of RLAs---i.e., reducing the sample size an RLA
requires when the reported outcome is correct---is an active field of research.
Efficient RLAs for \emph{instant-runoff voting}
(IRV), a common form of ranked-choice voting, were developed relatively
recently.  IRV is tallied in rounds.  The least popular candidate is eliminated
in each round until only one candidate remains: the winner.  In each round,
each ballot's most preferred choice among the remaining candidates is counted
as a vote for that candidate.  Tabulating an IRV election produces an
\emph{elimination order}; the last candidate in the order is the winner.

IRV is used in political contests in several countries: the federal House of
Representatives in Australia, along with most analogues at the state-level; the
president of India; single-winner contests in Ireland such as the president of
Ireland and by-elections to D\'ail \'Eireann (the Irish lower house); and
various contest in U.S.\ states including Alaska, California, Colorado, Maine,
and Nevada, and by some political parties for statewide primary elections.

RAIRE~\cite{blom2019raire} and AWAIRE~\cite{ek2023awaire} are the only extant
methods for conducting RLAs of IRV elections.  Both confirm the outcome by
ruling out all elimination orders that yield a different winner
(\emph{\altorder{s}}).  Both involve constructing `assertions' which, if true,
collectively rule out all \altorder{s}.\footnote{%
    The assertions used by RAIRE may also rule out some orders that correspond
    to the reported winner indeed winning, but through an elimination order
    that differs from the reported order.}
These assertions are then checked statistically using tools available in the
SHANGRLA framework for RLAs~\cite{shangrla,stark2023alpha}.  RAIRE is a
two-stage approach: generate a sufficient set of assertions before any sampling
(offline), then test the assertions by sampling.  AWAIRE perform both steps
simultaneously (online), using the sample to `learn' a sufficient set of
assertions that can be tested efficiently, while testing them.

Another difference between the two methods is that RAIRE requires \emph{cast
vote records} (CVRs, the voting equipment's internal record of the preferences
expressed on each ballot) to select the set of assertions to minimize expected
workload if the CVRs are accurate.\footnote{%
    If the CVRs are linked to the corresponding ballot cards, RAIRE and AWAIRE
    can use \emph{ballot-level comparison} auditing, increasing efficiency.
    See, e.g., Blom et al.~\cite{blomEtal20}.  ONEAudit~\cite{stark23ONEAudit}
    can also be used with RAIRE and AWAIRE to take advantage batch subtotals or
    linked CVRs.}

AWAIRE has benefits over RAIRE, leveraging the insight that one can decompose
every \altorder{} into a set of \emph{requirements} that must all be true for
the \altorder{} to be true, and let the sample dictate which requirements to
use to reject the \altorder{}, while, RAIRE pre-commits to a subset of
requirements to test.  Because of this, AWAIRE can adaptively identify the
requirements that are easiest to disprove, even when the CVRs are not accurate;
it does not require CVRs, but can use them if they are available; and it is
more resilient than RAIRE if the CVRs imply an incorrect elimination order but
the reported winner is correct.

Here, we address the main drawback of AWAIRE as presented so far: its
computational performance.  For contests with $\numcands$ candidates, the
original implementation of AWAIRE tracked and tested all $\numcands! -
(\numcands-1)!$ \altorder{s} and their numerous requirements.  That limited it
to elections with at most 6~candidates, fewer than many real-world IRV
elections.  We show how to vastly decrease the computational resources AWAIRE
needs.

The new implementation tracks a \emph{frontier} of suffixes of \altorder{s}.
Often, the sample allows AWAIRE to reject a suffix that an exponential number
of \altorder{s} share.  Otherwise, the new approach extends the suffix,
replacing it in the frontier with all suffixes with one more candidate.  As the
suffixes grow in length, they entail more requirements, which may make them
easier to reject.  By parsimoniously expanding suffixes, the algorithm never
needs to consider very many at a time, and most of the possible requirements
may never need to be tested.  We also consider forms of requirements that RAIRE
uses but that were not used in the original implementation of AWAIRE.  As a
result, the new version of AWAIRE is computationally tractable for IRV
elections with more than 50~candidates.


\section{Background}

We refer the reader to the original AWAIRE paper \cite{ek2023awaire} for
details of the notation and terminology, but we summarize the key objects and
ideas here: \emph{\altorder{s}}, \emph{requirements}, \emph{test
supermartingales} (TSMs), \emph{base TSMs}, and \emph{intersection TSMs}.

Let $\cand$ denote the set of candidates, $\numcands = |\cand|$ the number of
candidates, and $B$ the number of \emph{ballot cards}\footnote{%
In some countries, a ballot may comprise more than one piece of paper (card).}
cast in the contest.  We identify each ballot card $b$ with an ordering of a
subset of candidates, possibly the empty set.

An \emph{\altorder{}} is a candidate elimination order in which someone other
than the reported winner is last---i.e., the reported winner did not win.
There are $(\numcands-1)(\numcands-1)! = \numcands! - (\numcands-1)!$
\altorder{s}: for each of the $\numcands-1$ candidates who were not reported to
have won, there are $(\numcands-1)!$ elimination orders for the the other
$\numcands-1$ candidates that would make that candidate the winner.
\begin{example}
\label{ex:four}
Consider a four-candidate election, with candidates $\cw$, $\cx$, $\cy$, $\cz$,
where $\cw$ is the reported winner.  The outcome is confirmed if we can rule
out every elimination order in which any candidate other than $\cw$ is last
(every \emph{\altorder}):

{\centering{
$[\cw,\cx,\cy,\cz]$, $[\cw,\cx,\cz,\cy]$, $[\cw,\cy,\cx,\cz]$,
$[\cw,\cy,\cz,\cx]$, $[\cw,\cz,\cx,\cy]$, $[\cw,\cz,\cy,\cx]$, \\
$[\cx,\cw,\cy,\cz]$, $[\cx,\cw,\cz,\cy]$, $[\cx,\cy,\cw,\cz]$,
$[\cx,\cz,\cw,\cy]$, $[\cy,\cw,\cx,\cz]$, $[\cy,\cw,\cz,\cx]$, \\
$[\cy,\cx,\cw,\cz]$, $[\cy,\cz,\cw,\cx]$, $[\cz,\cw,\cx,\cy]$,
$[\cz,\cw,\cy,\cx]$, $[\cz,\cx,\cw,\cy]$, $[\cz,\cy,\cw,\cx]$. \\
}}

The other 6~elimination orders lead to $\cw$ winning: they are not
\altorder{s}.
\qed
\end{example}
Each \altorder{} is characterized by a set of \emph{requirements}, all of which
must be true for that \altorder{} to be the actual elimination order.  If any
requirement for an  \altorder{} fails, that rules out the \altorder{}---and
every other \altorder{} that shares that requirement.
AWAIRE used a single form of requirement:

\begin{description}
\item[Directly Beats.] $\DB(i, j, \stand)$ holds if candidate~$i$ has more
votes than candidate~$j$ when the only candidates remaining are $\stand
\supseteq \{i, j\}$: $i$ cannot be the next candidate eliminated when exactly
the candidates $\stand$ remain.
\end{description}
\noindent
If $\DB(i, j, \stand)$ is false, $j$ has more votes than $i$ when only $\stand$
remain standing.
In that case, $j$ cannot be the next eliminated: $i$ would be eliminated before
it.

Each requirement is associated with a \emph{test supermartingale} (TSM) called
a \emph{base TSM}.  (A TSM is a stochastic process starting at 1 that, if the
requirement is true, is a nonnegative supermartingale.  A nonnegative
supermartingale is like the fortune of a bettor in a series of games that are
fair or biased against the bettor, when the bettor is not allowed to go into
debt.  Ek et al.~\cite{ek2023awaire} explains how TSMs are used in AWAIRE.)

In turn, each \altorder{} is associated with a TSM called an \emph{intersection
TSM}, a weighted average of the base TSMs for the requirements that
characterize that \altorder{}.  If every requirement for that \altorder{} is
true, the intersection TSM for the \altorder{} is a nonnegative supermartingale
starting at 1.  The weights in the average are chosen adaptively to try to
minimize the sample size required to confirm the reported outcome (i.e., to
reject every \altorder{}) when the reported outcome is indeed correct.

Constructing the statistical tests of \altorder{s} from TSMs provides
\emph{sequential validity}: the evidence that the outcome is correct can be
evaluated after each sampled ballot card is examined, with no statistical
penalty for looking at the data repeatedly.  For any requirement that is true,
the chance that its base TSM ever reaches or exceeds $1/\alpha$ is at most
$\alpha$, by Ville's inequality.  If any \altorder{} is true, the chance that
its intersection TSM ever reaches the value $1/\alpha$ is at most $\alpha$.
Thus, if the audit stops without a full hand count only if every intersection
TSM hits or exceeds $1/\alpha$, the audit has risk limit $\alpha$.

Requirements are expressed in terms of the means of lists of numbers, one
number per ballot card (for each requirement).  An \emph{assorter} (see
Stark~\cite{shangrla}) assigns a number to each card, depending on the vote
preferences on the card and on the requirement.  The assorter assigns the
numbers in such a way that if its requirement is true, the mean of the list of
numbers is no larger than $1/2$.
\begin{example}
Consider the requirement $\DB(\cx, \cy, \cand)$ that candidate~$\cx$ beats
candidate~$\cy$ on first preferences.  The corresponding assorter assigns a
card the value $1$ if it shows a first preference for candidate~$\cy$, the
value $0$ if it shows a first preference for $\cx$, and the value $1/2$
otherwise.  If the mean of the resulting list of all $B$ numbers is less
than $1/2$, then the requirement $\DB(\cx, \cy, \cand)$ holds.
\qed
\end{example}
Each requirement can be tested by testing the statistical hypothesis that the
mean of its assorter values is at most $1/2$ from a random sample of values of
its assorter.  That is done by drawing ballot cards at random and computing the
value of the assorter corresponding to the requirement from the preferences on
each sampled card.  The same sample can be used to test all requirements by
computing the value of every assorter for each sampled card.

As in the previous implementation of AWAIRE, we use the ALPHA TSM with the
truncated shrinkage estimator to test requirements and intersections of
requirements from the sample of assorter values.  See
Stark~\cite{stark2023alpha} for more about ALPHA and Ek et
al.~\cite{ek2023awaire} for details on how ALPHA is used in AWAIRE.  Given the
cards sampled in draws $t=1, \dots, \ell$, let $(X_t)_{t=1}^\ell$ denote the
values assigned by the assorter of a particular requirement.  Let $M_\ell$ be
the TSM for the requirement, evaluated after the $\ell$th card is drawn.  It
can be written as a product:
\(
    M_\ell := \prod_{t=0}^{\ell} m_t.
\)
Here, $m_0 = 1$ is the initial value of the TSM and $m_t$ reflects the evidence
$X_t$ provides about the requirement: if $m_t > 1$, $X_t$ is evidence against
the requirement.
The TSMs for individual requirements are called \emph{base TSMs}.

To test an \altorder{} statistically, we could test each of its requirements
separately and reject the \altorder{} if we reject at least one of its
requirements.  However, doing this naively would increase the risk limit; this
is an instance of the well-known problem of \emph{multiple testing} in
statistics.  AWAIRE addresses multiple testing by forming a weighted average of
the base TSMs called an \emph{intersection TSM}, which is a nonnegative
supermartingale starting at 1 if every requirement for that \altorder{} is
true.

The weights are chosen \emph{predictably}: the weights at time $t$ depend on
the data collected up to time $t - 1$, but not on anything that has not been
observed before the $t$th card is drawn.  The intersection TSM is a product of
weighted means of the terms of the base TSMs.  When all the requirements hold,
the intersection TSM is a nonnegative supermartingale starting at 1.

Multiple weighting schemes were investigated in~\cite{voting24a}.  One of the
best and the simplest for AWAIRE was `Largest,' which puts all weight on the
base TSM that is largest at time $t-1$ (in the case of ties, it gives equal
weight to the largest).  We use `Largest' below because of its simplicity and
good empirical performance.

If every intersection TSM hits or exceeds $1/\alpha$, we can reject every
\altorder{}: the audit stops without a full hand count and the reported outcome
is certified.  Otherwise, AWAIRE continues until the sample contains every
ballot card: a full hand count.  The chance the audit stops without a full hand
count if any \altorder{} is correct is at most $\alpha$, the risk limit.

RAIRE avoids multiple testing by pre-commiting, before sampling commences, to a
sufficient set of requirements\footnote{%
    RAIRE uses the terminology `assertions' in place of `requirements'.}
that covers all \altorder{s}.  RAIRE uses the CVRs to select a set of
requirements that minimizes the expected number of ballot cards required to be
sampled to certify the contest, on the assumption that the CVRs are accurate.


\section{Improving AWAIRE}

The original implementation of AWAIRE tracked all $\numcands! - (\numcands-1)!$
\altorder{s} separately.  The requirements characterizing each \altorder{} were
all \DB{}.  Because there are so many \altorder{s}, the implementation became
computationally impractical for more than 6~candidates.

The present contribution makes AWAIRE tractable for contests with far more
candidates, using three tools: \emph{incremental expansion}, \emph{use of new
requirements}, and \emph{requirement abandonment}.  The first of these helps
the most, but for clarity we begin by describing the second.  The new
implementation of AWAIRE and the code and output for the figures and tables in
this paper are at \url{https://github.com/aekh/awaire}.

\subsection{Another Type of Requirement}

We introduce a new requirement to AWAIRE, related to a RAIRE assertion.
Candidate $i$ \emph{dominates} candidate $j$ if $i$ has more first-preference
votes than there are ballots that rank $j$ ahead of $i$ (including ballots that
mention $j$ but not $i$).  In other words, $i$ has more votes before any
candidate is eliminated than $j$ could ever possibly get, no matter who else is
eliminated.  The new type of requirement is the complement of this
condition:\footnote{%
    Blom et al.~\cite{blom2019raire} originally defined `WO' assertions, later
    renamed to `NEB'~\cite{blomEtal20}.  \DND{} is the negation of these.}
\begin{description}
\item[Does Not Dominate.]
$\DND(i, j)$ holds if candidate~$i$ does not dominate candidate~$j$: there
might be an elimination sequence that results in $j$ having more votes
than $i$.
\end{description}
If the requirement is false, $i$ dominates $j$: $j$ cannot possibly have more
votes than $i$.  The original implementation of AWAIRE used only $\DB{}$
requirements.  Including $\DND$ requirements can reduce sample sizes and
runtimes because the requirement $\DND(i,j)$ is shared by all \altorder{s} in
which candidate $i$ is eliminated before candidate $j$.  Although $\DND$
requirements may need larger samples to reject than $\DB$ requirements, they
still reduce runtime and there are only $\numcands (\numcands-1)$ of them.

Like $\DB$, the assorter for the requirement $\DND(\cx,\cy)$ assigns a ballot
card the value $0$ if it shows a first preference for candidate $\cx$ (so that
card will be counted for candidate $\cx$), the value $1$ if it shows a
preference for candidate $\cy$ before candidate $\cx$ or shows a preference for
candidate $\cy$ and does not mention candidate $\cx$ (so the card may
eventually contribute a vote to in $\cy$ before $\cx$ is eliminated), and the
value $1/2$ otherwise.

\subsection{Suffix Representation and Incremental Expansion}

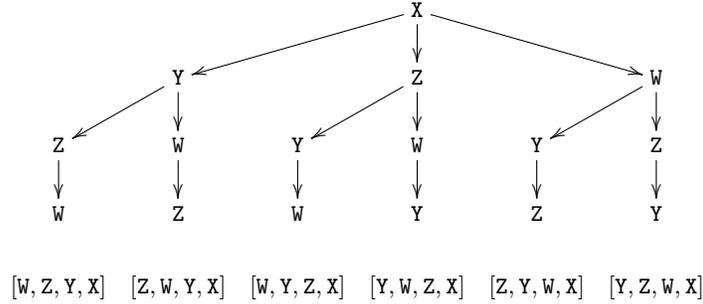
\begin{figure}[t]
\centering
\[
\xymatrix@C=1mm@R=5mm{
&&& \texttt{X} \ar[drr] \ar[d] \ar[dll] \\
& \texttt{Y} \ar[dl] \ar[d] &&
  \texttt{Z} \ar[dl] \ar[d] &&
  \texttt{W} \ar[dl] \ar[d] \\
\texttt{Z} \ar[d] & \texttt{W} \ar[d] &
\texttt{Y} \ar[d] & \texttt{W} \ar[d] &
\texttt{Y} \ar[d] & \texttt{Z} \ar[d] \\
\texttt{W} & \texttt{Z} &
\texttt{W} & \texttt{Y} &
\texttt{Z} & \texttt{Y} \\
 [\W,\Z,\Y,\X] & [\Z,\W,\Y,\X] & [\W,\Y,\Z,\X] &
 [\Y,\W,\Z,\X] & [\Z,\Y,\W,\X] & [\Y,\Z,\W,\X]
}
\]
\caption{The suffix tree for \altorder{s} with alternate winner \X.}
\label{fig:suffix}
\end{figure}

Instead of tracking all \altorder{s}, we track \emph{suffixes} of \altorder{s}.
Each suffix of a set of \altorder{s} can be represented by a set of
requirements that are shared by all \altorder{s} with that suffix.  As the
audit progresses, either it finds enough evidence to reject a suffix (along
with all \altorder{s} that include it), or it extends that suffix by one in all
possible ways and tests those extended suffixes.
\begin{example}
\label{ex:suffix}
Consider the \altorder{s} in \autoref{ex:four} where $\cx$ wins.
\autoref{fig:suffix} illustrates how these \altorder{s} can be represented
by a suffix tree.  On the first level of the tree, the suffix $[\dots,
\cx]$ encompasses all \altorder{s} where $\cx$ wins (listed at the bottom
of the tree).  One step below are three suffixes, $[\dots, \cy, \cx]$,
$[\dots, \cz, \cx]$, $[\dots, \cw, \cx]$, denoting the winner $\cx$ but
also the possible runner-ups ($\cy$, $\cz$, and $\cw$, respectively).  The
first of these represents the two complete \altorder{s} $[\W,\Z,\Y,\X]$ and
$[\Z,\W,\Y,\X]$.
\qed
\end{example}
Each suffix has an associated intersection TSM, a weighted combination of the
base TSMs for requirements shared by all \altorder{s} with that suffix.  If
that intersection TSM hits or exceeds $1/\alpha$, we reject every \altorder{}
with that suffix.

\begin{figure}[t]
\centering
\includegraphics[width=0.83\textwidth]{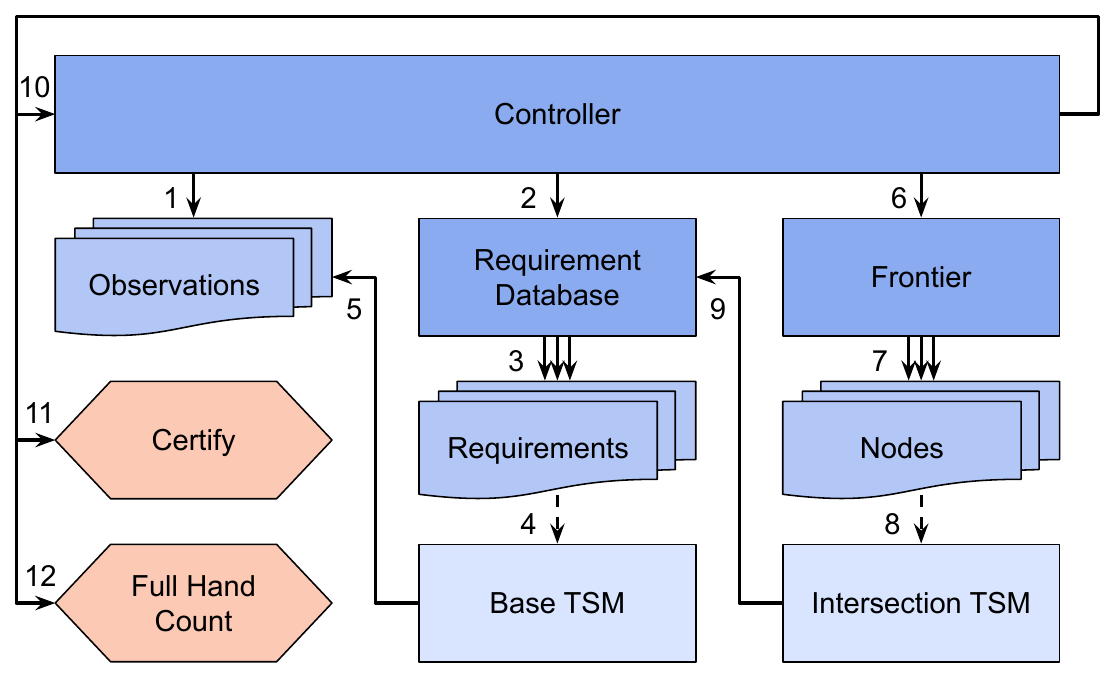}
\caption{\textbf{An overview of the new implementation of AWAIRE.}
The process begins by (1)~sampling some ballots at random.  Then, the
controller (2) prompts the requirement database, which (3) processes each
requirement in the database.  Each requirement has (4) a base TSM, the
value of which is (5) calculated from the ballots seen so far.  Once all
requirements have been processed, the controller (6) prompts the frontier,
which (7) processes and keeps track of all the nodes.  Each node represents
a suffix and has (8) an intersection TSM associated with it.  To calculate
the current value of the intersection TSM, it (9) requests the previous
values of the node's associated base TSMs (used as weights) and how their
values changed to reach the current values (used as returns).  If a node's
intersection TSM is above the risk threshold, we remove it; further, if a
node's intersection TSM and/or base TSM(s) are not increasing enough or at
all, the frontier may expand this node, introducing several children of
longer suffixes.  Once all nodes have been processed, AWAIRE either (10)
continues sampling if nodes and unseen ballots remain, (11) certifies the
election if no nodes remain, or (12) terminates as a full hand count if no
unseen ballots remain but nodes remain.  At any point in this process,
auditors may decide to perform a full hand count instead of continuing to
sample ballots at random; this cannot increase the risk.}
\label{fig:dataflow}
\end{figure}

The base TSM for each active requirement is only computed once and stored in a
dictionary (i.e., hash table).  The test for each suffix can access a set of
base TSM values and can determine weights to combine them into an intersection
TSM for that suffix.  The code can also remove from the database every
requirement that is no longer useful, further reducing memory and CPU usage.
\autoref{fig:dataflow} shows the structure of the algorithm.  The figure
caption summarizes the steps, many of which are described below in more detail.

\subsubsection{Suffix trees.}

The \altorder{s} of a given $\numcands$-candidate election can be represented
by a forest of $\numcands-1$ suffix trees, each rooted at the supposed winning
candidate according to that \altorder{}.  Each \altorder{} corresponds to the
unique path from a leaf node to a root node.  Recall the \altorder{s} defined
in \autoref{ex:four}.  The forest consists of 3~suffix trees rooted by the
candidates other than the reported winner: \texttt{X}, \texttt{Y} and
\texttt{Z}.  The one for $\cx$ is shown in \autoref{fig:suffix}; the other two
are analogous.  Each node in the forest corresponds to a suffix of
\altorder{s}.  For example, the node pictured under the root corresponds to
suffix $[\ldots, \cz,\cx]$ and subsumes (is the suffix of) the two \altorder{s}
$[\cw,\cy,\cz,\cx]$ and $[\cy,\cw,\cz,\cx]$.

\subsubsection{Node frontier.}

We keep track of a dynamic \emph{frontier} of nodes in this forest of suffixes,
the nodes for which we calculate intersection TSMs to test \altorder{s}.  Every
\altorder{} has a suffix in the frontier.  If we can rule out each node in a
frontier, we have ruled out every alternative winner of the
election.\footnote{%
    RAIRE also uses suffix trees, but it computes a \emph{static} frontier of
    the \altorder{} forest using the CVRs (before observing any sampled ballot
    cards).}

Before sampling commences, the frontier is initialized with $|\cand|-1$
suffixes of the form $[\ldots, c]$, for each candidate $c$ other than the
reported winner, i.e., the roots of the forest of \altorder{s}.  The
requirements for the root node labelled $c$ are $\{ \DND(c',c) \suchthat c' \in
\cand - \{c\} \}$, since, for $c$ to have won, no other candidate could have
dominated $c$.

\begin{example}
Continuing with \autoref{ex:four}, the requirements for the suffix [\ldots,\X]
are $\{ \DND(\Y,\X)$, $\DND(\Z,\X)$, $\DND(\W,\X) \}$.  The requirements
for the other suffixes [\ldots,\Y] and [\ldots,\Z] are similar.
\qed
\end{example}

Each node $m$ has a watchlist of requirements that are necessarily true if its
suffix is true.  These are the requirements that are shared among the
\altorder{s} represented by $m$.  The weighting scheme for $m$ is in essence no
different than for the original implementation of AWAIRE except it uses only
the base TSMs from the watchlist of requirements.  Thus, some requirements and
their base TSM values are ignored.
If the intersection TSM for node $m$ ever reaches $1/\alpha$, we can reject all
elimination orders with that suffix.  We remove node $m$ from the frontier, its
subtree is \emph{pruned}.

\subsubsection{Expanding nodes.}

If none of a node's requirements appears to be false (e.g., all the base TSMs
have decreased for a long time or are less than 1), we split that node.  Given
a node representing suffix $[\ldots, S]$ we split it to create the child nodes
$\{ [\ldots, c] \pp S \suchthat c \in \cand \setminus S \}$, where $\pp$
represents sequence concatenation.  Thus, we create a child node for each
candidate not appearing in the suffix $S$ of the expanded parent node.

Consider a suffix $[\ldots, c_\ell,c_{\ell-1},\ldots,c_1]$ with the unmentioned
candidates, implicitly eliminated before this suffix, represented by the set
$\mathcal U$.  The requirements of this suffix are given by:
$\{ \DND(c_j,c_i)\suchthat\allowbreak \ell \geqslant j > i \geqslant 1 \}$,
i.e., each candidate $c_j$ eliminated before $c_i$ \emph{does not dominate}
$c_i$; together with $\{ \DND(c,c_i) \suchthat \ell \geqslant i \geqslant 1, c
\in \mathcal U\}$, i.e., every unmentioned candidate $c$ \emph{does not
dominate} any candidate $c_i$ in the suffix; and $\{ \DB(c_i,c_j,\{c_\ell,
\dots, c_1\}) \suchthat \ell \geqslant j > i \geqslant 1\}$, i.e., just before
$c_j$ is eliminated, every other remaining candidate $c_i$ \emph{directly
beats} $c_j$.

Each node inherits all the requirements of its parent node, and adds more
specific requirements relating to the newly added candidate $c_{\ell+1}$.  We
only need to add the requirements
$\{ \DND(c,c_{\ell+1}) \suchthat c \in \mathcal U \setminus \{c_{\ell+1}\} \}$
and $\{ \DB(c_i,c_{\ell+1},\{c_{\ell+1}, \dots, c_1\}) \suchthat \allowbreak
\ell \geqslant i \geqslant 1 \}$
to the parent nodes requirements.

\begin{example}
Continuing our running example of \autoref{ex:four}, assume we decide to expand
the node $[\ldots,\X]$. We add three child nodes: $[\ldots,\Y,\X]$,
$[\ldots,\Z,\X]$, $[\ldots,\W,\X]$. The requirements for $[\ldots,\Y,\X]$
adds $\{ \DND(\Z,\Y), \DND(\W,\Y), \DB(\X,\Y,\{\X,\Y\}) \}$ to those
inherited from its parent $[\ldots,\X]$.
\qed
\end{example}

When a node is expanded, its intersection TSM (up to the latest sample) is
copied to all its children.  This step ensures that its continued use will
remain risk-limiting.

A critical ingredient for the improved AWAIRE is when and which nodes to
expand.  To decide \emph{which node} to expand we score nodes by the value of
its best performing base TSM. The higher the score the more likely we will be
able to reject this suffix.  So when we choose to expand a node we always
choose one with the lowest score.

To decide \emph{when} to expand a node we consider a few policies:
\begin{description}
\item[Every($i$).] We expand a node after every non-zero multiple of $i$
    ballots sampled.
\item[Below($x$).] After every ballot, we expand every node that has a score
    below $x$.
\end{description}
These policies are quite myopic, only looking at the current node's score.
We can also impose a look-ahead rule to avoid unnecessary expansions.
If we choose a node $m$ for potential expansion, we examine the child suffixes
of node $m$ and determine what their scores would be (by computing the base TSM
for any newly introduced requirements).
We only allow the expansion if:
\begin{description}
\item[Loose.] Some child node has a better score than $m$.
\item[Tight($y$).] Some child has a better score than $m$, and is also higher
    than $y$.
\end{description}

\subsection{Requirement Database and Requirement Abandonment}

The requirement database is a critical data structure of the algorithm as the
number of requirements is $\numcands (\numcands - 1) 2^{\numcands -
2}$.\footnote{%
    This is fewer than the number of \altorder{s} due to the \emph{order} of
    elimination being irrelevant for requirements; only the \emph{set} of
    eliminations is relevant.}
Thus, we have to aggressively restrict the number of requirements we track.

The requirement database is initially empty but nodes can \emph{request}
requirements needed for their intersection TSM, adding them to their watch-list
and the database (if not already added).  This happens when the frontier is
created or a node is expanded.  Adding a requirement to the database involves
calculating its base TSM from ballot card $1$ to the latest observed.

We can leverage some logical implications between requirements to decrease
computation time, by deciding to \emph{abandon} (i.e., set weight to 0 for the
remaining samples) particular requirements when there is sufficiently strong
evidence that they are true.  Note that this will not compromise the risk
limit, but it may increase the sample size required to terminate the audit (if
a requirement that is actually false is erroneously abandoned).
The two relationships we use are:
\begin{align*}
\neg\DB(i,j,S) \longleftrightarrow \DB(j,i,S) &&\text{and}&&
\neg\DB(i,j,S) \longrightarrow    \DND(i,j).
\end{align*}
Considering the requirement on the left-hand side of these rules, if its base
TSM exceeds $1/\alpha$ (our threshold of `enough evidence' that the requirement
is false\footnote{%
Due to multiple testing, this does not necessarily allow us to reject the
\altorder{s} it is part of.  To do that we need to use intersection TSMs.}),
we abandon the requirements on the right-hand side, since the evidence now
suggests that they ought to be true.

Another way a requirement can be abandoned is when it has been mathematically
proven to be true (i.e., we can show that the requirement must be true given
the number of remaining samples).

Finally, if, due to node pruning, a requirement is no longer part of any node's
watch-list, we need not process its base TSM.  In that case we \emph{park} the
requirement to save computation time.  If at another point this parked
requirement is requested, we simply \emph{unpark} it and calculate its base TSM
values from the time it was parked to the latest observed ballot.


\section{Analyses and Results}

\begin{table}[t]
\caption{Size of the final frontier for \AWAIREnew{}, showing the mean and 99th
percentile frontier size across all experiments on contests with a given number
of candidates. The second column shows the number of contests summarized in
each row, and the third column shows the total number of \altorder{s}
(max.\ possible frontier size).
The three subcolumns refer to different ways to specify $\eta_0$ in ALPHA:
either to 0.51, to the last-round margin (LRM), or to the reported assorter
margins (AM).}
\label{tab:froniersize}
\centering
\footnotesize
\begin{tabular}{rrrrrrrrr}
\toprule
 & & & \multicolumn{3}{c}{Mean} &
       \multicolumn{3}{c}{99th percentile} \\
\cmidrule(lr){4-6}
\cmidrule(l){7-9}
Candidates & Contests & \Altorder{s} &
                              0.51 &   LRM &    AM &   0.51 &    LRM &     AM \\
\midrule
 4 &  5 &     18 &               5 &     7 &     7 &     11 &     11 &     11 \\
 5 & 50 &     96 &               7 &     9 &    10 &     36 &     37 &     40 \\
 6 & 25 &    600 &              10 &    15 &    16 &     34 &     60 &     55 \\
 7 & 17 &  4,320 &              31 &    38 &    44 &    379 &    372 &    455 \\
 8 &  7 & 35,280 &              13 &    35 &    35 &     57 &    218 &    259 \\
11 &  2 & $4 \times 10^{7\phantom{0}}$
                           & 5,005 & 5,471 & 6,194 & 21,171 & 24,190 & 25,762 \\
18 &  1 & $6 \times 10^{15}$ &  17 &   937 &    81 &     17 & 22,711 &    694 \\
19 &  2 & $1 \times 10^{17}$ & 794 & 6,879 & 1,299 &  3,068 & 82,947 & 10,889 \\
36 &  1 & $4 \times 10^{41}$ & 170 & 3,463 &   740 &  1,318 & 51,669 &  5,075 \\
\bottomrule
\end{tabular}
\end{table}

We used the data from the 93~New South Wales Legislative Assembly Contests and
14~contests in the USA used by \cite{blom2019raire}.\footnote{%
    \url{https://github.com/michelleblom/margin-irv/} (visited 16~May 2024).}
We also used datasets for three contests for Minneapolis Mayor (in 2013, 2017,
and 2021),\footnote{%
    \url{https://vote.minneapolismn.gov/results-data/election-results/}
    (16~May 2024).}
for a total of 110~contests.

The \emph{reported margin} (in cards) of an election is the minimum number of
cards that must have been mistabulated if the reported winner really lost.  We
use \emph{margin} to mean \emph{reported diluted margin}, the reported margin
in cards divided by the number of cards from which the sample is drawn.  We
used \texttt{margin-irv}~\cite{blom2018margin} to find margins for 109 of the
contests; it did not find the margin for 2021 Minneapolis Mayor (19~candidates)
in a week.

When the reported outcome is correct, audit sample sizes can generally be
reduced by exploiting information about the tabulation available before
auditing, for instance, the \emph{reported last-round margin}.  Often, the
reported last-round margin is close to or equal to the actual reported margin
in cards.\footnote{%
    Of the 109 contests for which we calculated the margin, 8 had last-round
    margins greater than their actual margin.  The difference ranged from $11$
    to $2,539$ ballots, equating up to a few percentage points in margin
    relative to the total number of ballots.}

\begin{figure}[t]
\centering
\includegraphics[width=\textwidth]{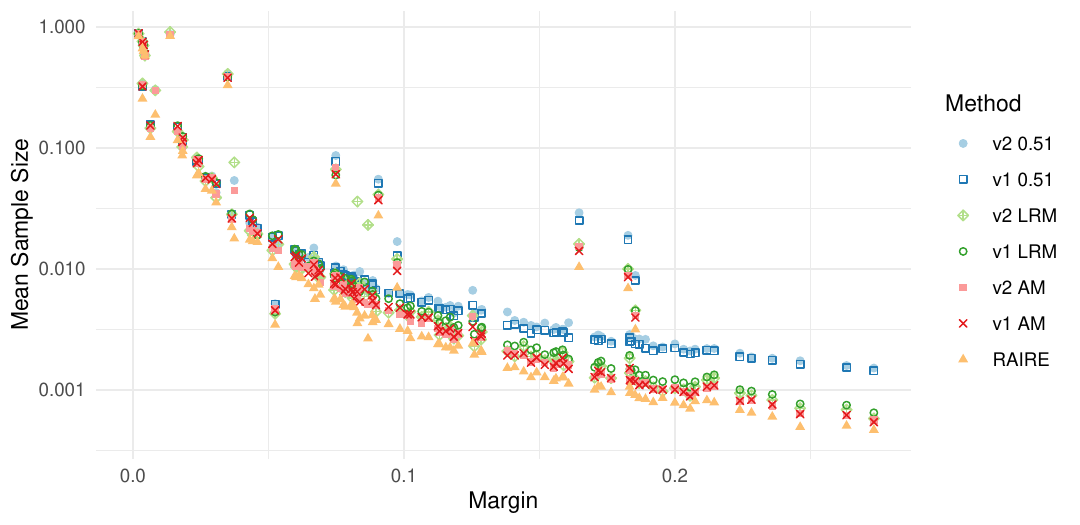}
\caption{Mean sample size (as a fraction of the total number of ballots)
comparing \AWAIREnew, \AWAIREold, and RAIRE at risk limit $0.05$.  All used the
ALPHA test supermartingale with $d = 200$ in the truncated shrinkage estimator
and $\eta_0 = 0.51$, the last-round margin (LRM), or the reported assorter
margins (AM).  We omitted the San Francisco Mayor 2007 contest (the margin was
much larger than for the other contests), but see \autoref{tab:handpicked}.}
\label{fig:sample}
\end{figure}

We simulated 500 ballot-polling audits for every contest, with each audit
corresponding to a randomly sampled (without replacement) order of the ballots.
The same 500 sampled orders for each contest were used across all methods.  The
ballots were selected one at a time.  After each ballot, the method under
experiment was used to determine whether to terminate and certify the contest
(with risk limit $\alpha = 0.05$), or continue sampling.

We compared the old and new implementations of AWAIRE (v1 and v2, respectively)
and RAIRE.  Each simulation had access to 32GB of RAM.  For the tuning
parameters in the truncated shrinkage estimator for ALPHA, we used $d = 200$
and three choices of $\eta_0$:
\begin{itemize}
\item \emph{0.51}: setting $\eta_0 = 0.51$, as recommended by~\cite{voting24a}
\item \emph{LRM}: setting $\eta_0$ for all requirements using the last-round margin
\item \emph{AM}: setting $\eta_0$ to the reported assorter margin (for each
    requirement separately), which requires CVRs.
\end{itemize}
All experiments with RAIRE used AM.

In earlier extensive comparisons of expansion schemes in \AWAIREnew{},
\textbf{Below}($1$)--\textbf{Tight}($e^{0.5}$) consistently performed the best.
We used this scheme for all \AWAIREnew{} experiments reported here.  Additional
experiments with and without requirement abandonment and \DND{s} showed some
performance improvements (without affecting sample sizes) when using the above
expansion scheme.  We have omitted the details due to space constraints.

\subsection{Computational Performance}

Incremental expansion lets the audit `group reject' many \altorder{s} by
rejecting nodes they share, rather than having to reject all
$\numcands!-(\numcands-1)!$ \altorder{s} separately.  One measure of the
computation saved is the final frontier size (the number of nodes that were not
expanded but instead pruned) compared to $\numcands!-(\numcands-1)!$, the total
number of \altorder{s} and thus the maximum possible frontier size; see
\autoref{tab:froniersize}.  Incremental expansion saves an exponential amount
of memory.\footnote{%
    The result for 18~candidates represents a single contest (San Francisco
    Mayor 2007) that was inexpensive to audit.  There is little expansion with
    0.51 and AM but quite a bit with LRM.  Nonetheless, with LRM, the audit
    terminated after 24~ballots on average, compared to 60 for 0.51, since LRM
    expanded to nodes that were easy to reject.  Using AM expanded to fewer
    nodes but on average terminated after 20~ballots.}

\begin{figure}[t]
\centering
\includegraphics[width=\textwidth]{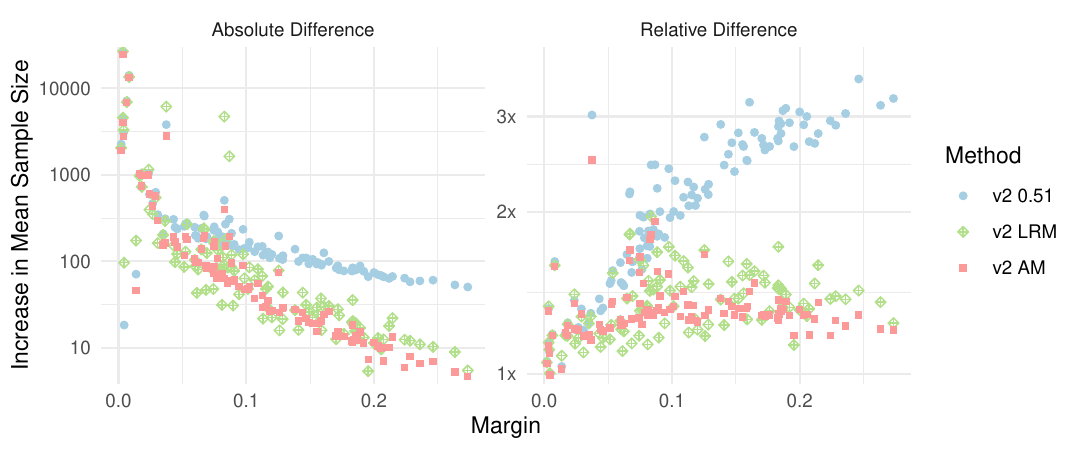}

\caption{Average number of cards sampled by \AWAIREnew{} minus average number
sampled by RAIRE, as a function of contest margin, to audit 109~contests at
risk limit 0.05.
Omitted:
Lismore (left pane, \AWAIREnew{} AM samples 91 fewer cards than RAIRE on
    average),
Minneapolis Mayor 2013 using LRM (right pane, relative differences are beyond
    4$\times$), and
San Francisco Mayor 2007 (both panes, margin too large).}
\label{fig:sizediff}
\end{figure}

\AWAIREnew{} was substantially faster, scaling exponentially better in
$\numcands$.  For elections with 4--8 candidates it saved seconds for the
smaller elections and up to 20 minutes on the larger elections.  \AWAIREold{}
could not complete the audit of any contest with more than 8~candidates (6~of
the contests) regardless of the margin of victory, for lack of memory.
\AWAIREnew{} could complete all but 36~simulated audits (two using $\eta=0.51$,
33 using LRM, and one using AM; all for the Minneapolis contests) out of
165,000, for lack of memory or time (48 hours).  This could be resolved by
further experimentation with expansion schemes.  We treated those 36~audits as
full hand counts.

To stress-test the implementation, we added `fake' candidates to a handful of
contests.  These candidates never get any votes, but the audit cannot foresee
that, so it must include them in the search tree.  The new implementation could
easily handle 55~candidates in those simulations, and possibly more.  The
runtime was always within a minute \emph{per ballot} on average, and only
reached an hour \emph{per audit} on average in the toughest cases.  The largest
real IRV election to the authors' knowledge had 36~candidates (Minneapolis
Mayor 2013).  CPU time per audit for RAIRE and our implementation of
\AWAIREnew{} were similar.

\subsection{Statistical Performance}

To quantify the statistical efficiency we used the sample size required to
certify each contest, averaged across simulated audits.

The mean sample size as a proportion of the total number of ballots is shown in
\autoref{fig:sample}.  Unsurprisingly, RAIRE is typically the most efficient
since it uses CVRs, but \AWAIREnew{} is close or on par (for Lismore).  Having
more information (LRM) is better than default (0.51) for both \AWAIREold{} and
v2; v2 was slightly better than v1.  The mean sample size for \AWAIREnew{} was
never more than v1 by more than 1.8\% of the total number ballots or 55~ballots
(despite having less information at the start), and was often slightly more
efficient (likely due to the difference in initial bets).

\begin{table}[tb]
\caption{Contest sizes, no.\ candidates, margins, and audit sample sizes for
8~contests.
Rows 5--11: mean sample size for AWAIRE with various settings and RAIRE. \\
Bottom row: largest standard error of the mean sample sizes in each column.}
\label{tab:handpicked}
\centering\footnotesize
\begin{tabular}{rlrrrrrrrr}
\toprule
\multicolumn{2}{l}{\textbf{Contest:}} &
\hspace{-5cm}\rotatebox[origin=r]{-20}{Lismore} &
\hspace{-5cm}\rotatebox[origin=r]{-20}{Aspen City Council} &
\hspace{-5cm}\rotatebox[origin=r]{-20}{Monaro} &
\hspace{-5cm}\rotatebox[origin=r]{-20}{Auburn} &
\hspace{-5cm}\rotatebox[origin=r]{-20}{Macquarie Fields} &
\hspace{-5cm}\rotatebox[origin=r]{-20}{Maroubra} &
\hspace{-5cm}\rotatebox[origin=r]{-20}{Cessnock} &
\hspace{-5cm}\rotatebox[origin=r]{-20}{San Francisco Mayor} \\
\multicolumn{2}{l}{\textbf{Candidates:}} & 6 & 11 & 5 & 6 & 7 & 5 & 5 & 18 \\
\multicolumn{2}{l}{\textbf{Margin:}}  & 0.44\% & 1.38\% & 2.43\% & 5.15\%  &
                                        7.43\% & 10.1\% & 20.0\% & 34.0\%  \\
\multicolumn{2}{l}{\textbf{Ballots:}} & 47,208 &  2,544 & 46,236 &  44,011 &
                                        47,381 & 46,533 & 45,942 & 149,465 \\
\midrule
\multirow{6}*{\rotatebox[origin=c]{90}{AWAIRE}}
& v1 0.51 & 28,088 & ---   & 3,642 & 803 & 438 & 264 & 102 & --- \\
& v1 LRM  & 28,596 & ---   & 3,758 & 826 & 420 & 222 &  56 & --- \\
& v1 AM   & 27,851 & ---   & 3,660 & 709 & 357 & 196 &  46 & --- \\
& v2 0.51 & 27,204 & 2,200 & 3,446 & 794 & 440 & 286 & 110 & 60  \\
& v2 LRM  & 27,282 & 2,303 & 3,245 & 626 & 317 & 191 &  47 & 24  \\
& v2 AM   & 27,095 & 2,175 & 3,453 & 656 & 342 & 191 &  48 & 20  \\
\multicolumn{2}{l}{RAIRE} &
            27,186 & 2,129 & 2,850 & 539 & 269 & 143 &  36 & 16  \\
\addlinespace
\multicolumn{2}{l}{\textit{Std.\ err.\ (max.)}} &
             410.0 &   8.5 & 112.9 & 27.5 & 13.4 & 7.9 & 1.8 & 0.7 \\
\bottomrule
\end{tabular}
\end{table}

For larger elections we can only compare RAIRE to \AWAIREnew{}.
\autoref{fig:sizediff} shows the absolute and relative increase in mean sample
sizes for RAIRE (using error-free CVRs) and the new implementation.  While the
difference in the number of cards can be large for large elections with tiny
margins, the relative difference is small; and while the relative difference is
large for small elections, the difference in cards is small.

\autoref{tab:handpicked} shows detailed results for a few elections with
various margins.  The narrower the margin, the better RAIRE typically does
(since it takes advantage of the accurate CVRs), but the relative difference is
small.  For the new implementation, using the LRM usually helps, but using
individual assorter margins did not help more.

\begin{table}[t]
\caption{Average number of ballots required to certify the winner in the
Strathfield (top, 46,644 cards cast) and Ballina (bottom, 47,865 cards cast)
contests at risk limit $0.05$ when the candidates are re-labeled, for RAIRE and
the new implementation of AWAIRE (v2) with different ways of choosing $\eta_0$.
`F' means the audit led to a full hand count in every run.
Notation for reported elimination orders:
an integer means the candidate with that number is in that place in the order;
a crossed-out integer means the given candidate is not in that place;
and a dot ($\cdot$) means any unmentioned candidate can be in that place.
The final row is the only order with a different winner.  Ranges span the
lowest and highest mean sample size of all permutations of a row.}
\label{tab:permuted-candidates}
\centering
\footnotesize
\begin{tabular}[t]{ll@{}l@{}l@{}l@{}l@{}l@{}lrrrr}
\toprule
\multicolumn{8}{l}{\textbf{Method:}}   & \multicolumn{3}{c}{\AWAIREnew}
                                       & \multirow{2}*{RAIRE} \\
\cmidrule(lr){9-11}
\multicolumn{8}{l}{\textbf{Reported}}  & 0.51 & LRM & AM &  \\
\midrule
\multirow{6}*{\rotatebox[origin=c]{90}{\textit{Strathfield}}}
&~&~&1&2&3&4&5                    & 6,491 & 6,495 & 6,553          &  5,626 \\
&~&~&$\cdot$&$\cdot$&$\cancel{3}$&4&5 & 6,491 & 6,495 & 6,553      &  5,626 \\
&~&~&$\cdot$&$\cdot$&4&$\cdot$&5  & 6,491 & 6,495 & 17,125--17,183 & 45,945 \\
&~&~&$\cdot$&4&$\cdot$&$\cdot$&5  & 6,491 & 6,495 & 20,642--20,681 & 45,014 \\
&~&~&4&$\cdot$&$\cdot$&$\cdot$&5  & 6,491 & 6,495 & 20,848--20,879 & 45,014 \\
&~&~&1&2&3&5&4                    & F     & F     & F              & F      \\
\midrule
\multirow{9}*{\rotatebox[origin=c]{90}{\textit{Ballina}}}
&1&2&3&4&5&6&7                               & 3,777 & 3,836 & 3,707        & 2,737\\
&$\cdot$&$\cdot$&$\cdot$&$\cancel{4}$&5&6&7  & 3,777 & 3,836 & 3,682--3,730 & 2,737\\
&$\cdot$&$\cdot$&$\cdot$&$\cdot$&6&$\cdot$&7 & 3,777 & 3,836 & 3,802--4,598 & 47,422--F\\
&$\cdot$&$\cdot$&$\cdot$&$\cdot$&$\cancel{5}$&6&7 & 3,777 & 3,836 & 4,409--5,556 & 47,422--F\\
&$\cdot$&$\cdot$&$\cdot$&6&$\cdot$&$\cdot$&7 & 3,777 & 3,836 & 5,323--6,320 & F\\
&$\cdot$&$\cdot$&6&$\cdot$&$\cdot$&$\cdot$&7 & 3,777 & 3,836 & 6,478--7,203 & F\\
&$\cdot$&6&$\cdot$&$\cdot$&$\cdot$&$\cdot$&7 & 3,777 & 3,836 & 8,019--8,439 & F\\
&6&$\cdot$&$\cdot$&$\cdot$&$\cdot$&$\cdot$&7 & 3,777 & 3,836 & 7,876--8,307 & F\\
&1&2&3&4&5&7&6                               & F     & F     & F            & F\\
\bottomrule
\end{tabular}
\end{table}

\subsection{Robustness to CVR errors}

Ek et al.~\cite{ek2023awaire} illustrated the robustness of \AWAIREold{}
compared to RAIRE when the CVRs have errors.  We repeated that experiment using
\AWAIREnew{} for the Strathfield and Ballina contests; see
\autoref{tab:permuted-candidates}.  For each contest we re-labelled the
candidates on the ballots and ran 200 simulated audits for each re-labeling and
method.  In all renumberings but the last, the winner is unchanged.  The
workload for approaches that do not use (erroneous) information to set assorter
margins was unchanged by renumberings that do not change the winner.  RAIRE
become much worse as the CVRs increasingly became less accurate; \AWAIREnew{}
using AM was affected less.  For Ballina, RAIRE often led to an unnecessary
full hand count.


\section{Discussion}

The new implementation of AWAIRE (v2) has comparable statistical efficiency to
the original (v1) but requires substantially lower computational resources,
allowing audits of IRV elections with up to 55~candidates.  Using an
incremental expansion strategy for AWAIRE does not undermine its risk-limiting
properties.  It amounts to giving zero weight to the base TSMs for a subset of
requirements for a group of \altorder{s}.  Expanding the frontier is equivalent
to changing the weights from zero to something positive, using past samples to
inform the choice of weights.  Because only past samples are used to select the
weights, the stochastic process is still a TSM.

Future work includes understanding how to better leverage CVRs in AWAIRE when
using incremental expansion, e.g., how to `pre-expand' the frontier, perhaps
guided by RAIRE-produced assertions; using AWAIRE for comparison audits
including those based on assorter means for groups of CVRs
\cite{stark23ONEAudit}; experimenting with expansion strategies and other
weighting schemes and ALPHA tuning parameters; and experimenting with more
varieties of CVR errors.


\bibliographystyle{splncs04}
\bibliography{references}


\end{document}